# Decrease of Particle Interference Pattern Due to Energy Changes.

Erez .M. Yahalomi

Raanana,Israel.
Email:erezsu@hotmail.com

In this paper we reveal an additional new cause for decreasing of particles interference. We found that particle which, transmits kinetic energy to the detector decrease its wave packet area size. A detector that transmits kinetic energy to the particle increases the particle wave packet area. We found these phenomena cause decrease in the interference pattern .We show that the interference pattern decrease is graduate and proportional to the amount of the particle kinetic energy change. We show that our findings are in agreement with the results of different experiments like the two slits experiment and the two paths interferometer experiment. We conclude the wave like behavior of the particle, does not have to lost in the detection process, it only changes the particle area size.

Recent development in the experimental technology enables to study quantum mechanics at the level of a single particle.For instance, experimental realization of the two slits experiment [1,2] .In this experiment a single particle can propagate through two slits and exhibits a self interference characteristic. The particle interference patterns is loss when a particle location detector is activated.
Additional experiment is the two path interferometer experiment wherein an electron can propagate through two separate channels made of with joint beginning and ending. Electron interference effect is obtained at the joint ending of the two channels. Once a detector, which is located in proximity to one of the channels, is activated a reduction of the interference visibility is obtained. The experiment is explained by dephasing of the electron [3,4] .

Analyzing the two slits and two paths experiments we revealed a new additional cause for the loss of particle interference pattern. In this paper we show that the particle wave function size changes, due to energy exchange with a detector. Causing decreasing in the interference pattern.

We describes the particle wavefuncion as a wave packet [6,7]. Looking at the detection operation in the two slits experiment, we show that a particle observed by a detector that absorbs kinetic energy from the particle, result in decrease of the particle wave function size. This detector can be for example, a photo electric detector or some types of electronic detectors. We show that an increase in the particle kinetic energy which can be caused by a second type of detector increase the wave packet size.

There are various kinds of wave packets that are appropriate solution of the Schrodinger equation. These wave packet differ by the weight function A(k) . Equation.1 is a general equation for a one dimensional wave packet,

$$\psi(x,t) = \frac{1}{\sqrt{2\pi}} \int_{-\infty}^{\infty} e^{i(kx - \hbar k^2 t / 2m)} A(k) \, dk . \qquad [1]$$

For the wave packet function we chose Gottfried`s [7] function, eq.2 . We choose this function because it has small dispersion .

$$A(k) = \frac{1}{(k - k_o + i\kappa)^n} . \qquad [2]$$

$k_0$ is the mean wave vector value. The momentum uncertainty is of order of $\kappa\hbar$. Gottfried wave packet dispersion is small if $t\Delta p/m \ll \Delta x_{t=0}$, where t is the propagation time, m is the particle mass, $\Delta p$ is the particle wave function momentum uncertainty and $\Delta x_{t=0}$ is the wave function position uncertainty at t=0.

We now obtain a three dimensional solution to the problem of reduction of the particle wave packet size distribution in space due to loss of kinetic energy.
The Schrodinger`s equation in spherical coordinates [9],

[3]

$$\frac{\hbar^2}{2m}\left[-\frac{1}{r^2}\frac{\partial}{\partial r}\left(r^2 \frac{\partial \psi}{\partial r}\right) + \frac{L^2}{r^2}\psi\right] + U(r) = E\psi$$

where r is the radius vector. L is the angular momentum U( r ) is the potential. The solution for a free particle ,U( r )=0 with a radial dependence only,l=0. Is a product of the angular part $Y_{00}=(1/4\pi)^{0.5}$ and a radial part given by Hankel functions, we choose the stationary spherical outgoing Hankel function

$$R^+_{kl} = \sqrt{1/4\pi}\, \frac{e^{ikr}}{r}$$

The index k represents the various eigenvalues associated with the same value of l. $R_{kl}+$ is Hankel function of the first kind. $N_k$ is a normalizaton constant equals, $N_K=(m/k\hbar)^{0.5}$.

Multiplying the two solutions parts we get the isotropic $\psi_{k00}$ wave function

[4]

$$\psi_{k00} = \sqrt{1/4\pi}\, \frac{N_k e^{ikr}}{r}$$

Equation 4 is a solution of the Schrodinger function for free particle in spherical coordinates. Inserting the free wave function into a potential step cause loss of the wave function kinetic energy at an amount of the potential energy step. On condition the free wave function`s initial kinetic energy is higher than the step potential. We choose the constant radial potential $U =U_0$ for r > 0, this potential can be for example a coulomb potential between an electron and a proton the attraction potential slows the electron movement and the electron loss kinetic energy. Different values of the potential represent different values of kinetic energy, our aim is to show correlation between the particle (in this example the electron) amount of kinetic energy loss and decrease in the particle wavefuction size. The constant radial potential can be for instance a coulomb potential act on a distance larger than the electron wave packet size, in this case the coulomb potential can be approximated as constant in all the wave packet area. The constant potential represents the net loss of the particle kinetic energy.

Substituting the potential in eq.1 we obtain the outgoing wave function solution.

$$\psi_{k00} = \sqrt{1/4\pi} N \frac{e^{i\frac{1}{\hbar}\sqrt{2m(E-U0)}r}}{r}$$

$$k = \sqrt{\frac{2mE}{\hbar^2}}$$

Where ,N is a normalization constant. E is the kinetic energy of the free wave function. We define the term $K_0$,

[5]

$$K_0 \equiv \sqrt{\frac{2mU_0}{\hbar^2}} = \sqrt{\frac{2m\Delta E}{\hbar^2}}$$

where $\Delta E$ is the kinetic energy reduced from the particle.

Assuming the main wavevectors region is peaked around $k_0$, Taylor approximation around $k_0$ is,

$$\sqrt{k^2 - K_0^2} \approx \sqrt{k_0^2 - K_0^2} + (k - k_0)\left[\frac{d\sqrt{k^2 - K_0^2}}{dk}\right]_{k=k_0}$$

$$\approx q_0 + (k - k_0)\frac{k_0}{q_0}$$

where,

$$q_0 = \sqrt{k_0^2 - K_0^2}$$

We describe the particle, as a wave packet composed of spherically symmetric stationary eigenfunctions of the spherical Schrodinger equation. The free particle wave packet is,

[6]

$$\Psi_i = \frac{N_i}{\sqrt{2\pi}}\sqrt{1/4\pi}\frac{e^{ik0r}}{r}\int_0^\infty \frac{I}{(k - k_0 + i\kappa)^n}e^{i(k-k_O)r}dk$$

If κ<<$k_0$ and n>>1 this wave packet is sharply peaked in momentum space about the mean value $k_0$. $N_i$ is a normalization constant, which depends on the values denoted to κ,$k_0$ and n. Since we chose the spherical wavepacket as a combination of out going wave function solutions the integration range in eq.6 is on wavevectors values k >0. The particle wave packet after the particle loss of kinetic energy is a superposition of the solutions of the spherical Schrodinger equation with a potential step,

$$\Psi_t = \frac{N_t}{\sqrt{2\pi}}\sqrt{1/4\pi}\frac{e^{iq_0 r}}{r}\int_0^\infty \frac{1}{(k-k_0+i\kappa)^n}e^{i(k-k_0)\frac{k_0}{q_0}r}dk$$

[7]

We approximate the weight function A (k) as unchanged after the kinetic energy change. Because the main wavevector region, which is the wavevectors region with high probability coefficients is narrow comparing to the all wavevectors region. Additional reason is although A(k) is not changed after the energy change. Each A(k) coefficients relate after the energy change to a different wavevector than the wavevector before the change. Instead of the wave vector k, after the kinetic energy loss A(k) is the weight coefficient of the wavevector $(k^2-K_0^2)^{0.5}$.
We get proportional distribution weight coefficient for a different wavevectors region.  The weight function indicate that most of wavevectors are distributed in a small region around $k_0$, the probability coefficient of the wave vectors outside this region is negligible. In the main wavevectors region the kinetic energy of the wavevectors is larger than $U_0$ and the transmission coefficient is close to unity. We then approximated the transmission coefficient of $\psi_t$ as 1 and the reflected part is negligible. We write equation .7 in terms of equation.6,

$$\psi_t(r) = N'\psi_i(\frac{k_0}{q_0}r)$$

where N`=$N_t/N_i$ is the ratio between the normalization constant of $\psi_t$ and $\psi_i$ respectively.  The relation between the wave function volume after it loss kinetic energy $V_t$ and the free wave function volume $V_i$ is,

[8]

$$V_t = \frac{4}{3}\pi \left(\frac{q_0}{k_0}\Delta r_i\right)^3 = \left(\frac{q_0}{k_0}\right)^3 V_i$$

Where $\Delta r_i$ is the free wave packet radius.

In figure.1. we calculate the wave function volume from eq.8 for different loss values of kinetic energy. The initial kinetic energy is 3ev, the initial wave function radius vector uncertainty is 0.79 μm, the initial wave function`s volume is 0.26 (μm)^3.

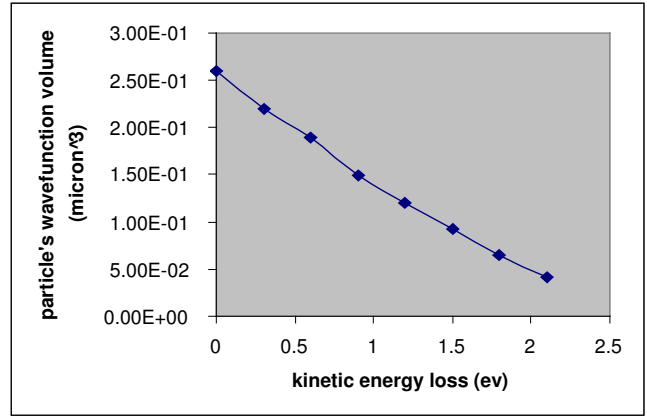

Figure .1. Volume distribution of particle spherical wave function

as a function of the kinetic energy the particle loss.

The kinetic energy decrease of the wave function is expressed by a change in the wavevector value as seen in eq.7 The width of a wave packet is determined by interference between the wavevectors of the wave packet. Let the main range of the wavevectors in the wave packet be $[k_1,k_2]$. When the particle loss kinetic energy as in various detection processes the wave vectors are changing. Kinetic energy decrease of $\Delta E$ will change the wave vector k to $(k^2-K_0^2)^{0.5}$ where $K_0$ is defined in eq.5. The main wavevectors range changes to
$[(k_1^2-K_0^2)^{0.5},(k_2^2-K_0^2)^{0.5}]$ the main region after the decrease in the particle kinetic energy is larger than $[k_1,k_2]$ a broader range of wavevectors influencing the interference in the wave packet result in decrease of the wave packet volume size.

We now look at the dynamics of the process. We discuss the one dimension case but it can extent to two and three dimensions. Let a free electron wave packet propagate in the x direction. The electron total energy E is, $E = \langle E_k \rangle + \langle U \rangle$. Where $E_k$ is the electron kinetic energy and U is the electron potential energy. The electron wave packet encounter a potential step of value $U_0$, We take $E > U_0$. Transmission through the potential step causes the particle to lose kinetic energy at the amount of the potential value $U_0$, $U_0 = \Delta E_k$. The change in the energy is shown in figure.2, figure.2a shows the potential of a free propagating electron one-dimensional wave packet then reaching a potential step and transmitted through it, we assume the reflected part is negligible. Fig .2b is the graph of the kinetic energy of the electron, the electron has a constant value until it transmits through the potential step then the electron`s kinetic energy is reduced in an amount equal to the potential energy $U_0$. A potential step constant in space and time is similar to irreversible loss of kinetic energy. Then the Schrodinger equation in a potential step is appropriate for describing an irreversible kinetic energy loss. At least for cases when the kinetic energy before transmitted through the potential step is higher than the potential step.

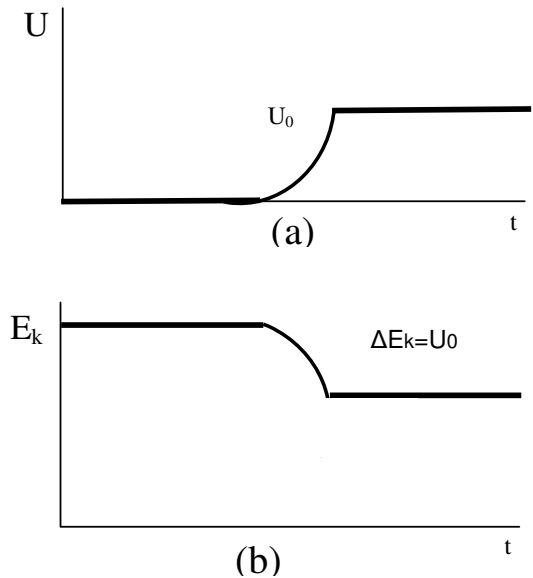

Fig.2. Energy dynamics of electron with potential step: (a) electron potential energy before and after encountering a potential step; (b) The electron`s kinetic energy variation in the process.

To demonstrate the effect of a particle detection we look at the two slits experiment[1,2], in the experiment a particle that propagates through two microscopic slits, exhibits an interference pattern of a particle wave function which pass through the two slits and interferes with itself. When the particle interact with a detector which detect through which slit the particle propagate the interference pattern decreases. Our finding is that kinetic energy is transmitted in irreversible way from the particle to the detector causing the particle wave function area to decrease. The decrease in the particle wave function in the axis parallel to the line connecting the two slits cause the wave function to be narrow enough to pass through only one slit. For this case we solve the two dimensional time dependent Schrodinger`s equation ,

$$i\hbar \frac{\partial \psi(x,y,t)}{\partial t} = \frac{-\hbar^2}{2m}\frac{\partial^2 \psi(x,y,t)}{\partial x^2} + \frac{-\hbar^2}{2m}\frac{\partial^2 \psi(x,y,t)}{\partial y^2} + U\psi(x,y,t)$$

[9]

The x dimension represents the particle wave packet dimension along the propagation direction, the y dimension represent the particle wave packet dimension parallel to the line along the two slits.
The free particle wave packet solution, where U=0, is

[10]

$$\psi_i(x,y,t) = \frac{n_i}{2\pi}\int A(k_x)A(k_y)e^{i(k_x x + k_y y) - \frac{\hbar}{2m}(k_x^2 + k_y^2)t} dk_x dk_y$$

$k_x$ and $k_y$ are the wave vectors on x and y axis respectively. $n_t$ and $n_i$ are the normalization constants of $\psi_t$ and $\psi_i$ respectively.
The weight functions $A(k_x), A(k_y)$ are in the x and y axes respectively.

$$A(k_x) = \frac{1}{(k_x - k_{ox} + i\kappa_1)^n} \quad ;$$

$$A(k_y) = \frac{1}{(k_y - k_{oy} + i\kappa_2)^n} \quad .$$

The electron penetrate a two dimensional potential step $U(x,y)=U_1(x)+U_2(y)$. $U_1(x)$ is the potential in x direction $U_2(y)$ is the potential in y direction. $U_1(x)=U_x$ for x>a  $U_2(y)=U_y$ for y>b a, b, $U_x$ and $U_y$ are constants. The physical meaning of this potential can be interpreted as interaction of the electron with electronic detector where at a certain distance the coulomb potential can be approximated as constant in the area of the electron wave packet. The alternative is that, the potential represents the energy of a photon emitted by the observed electron. Substituting these potentials in Schrodinger`s equation we obtain the solution,

[11]

$$\psi_t(x,y,t) = \frac{n_t}{2\pi} \int A(k_x) A(k_y) e^{i(\sqrt{k_x^2 - K_{0x}^2})x - \frac{\hbar}{2m}(k_x^2 - K_{0x}^2)t + \sqrt{(k_y^2 - K_{0y}^2)} y - \frac{\hbar}{2m}(k_y^2 - K_{0y}^2)t} dk_x dk_y$$

For $E > U_0$ the transmission coefficient can be approximated by one. We expand around $k_{0x}$ and $k_{0y}$ in a similar way to eq.7,
For the time evolution factor we do other expansion,
$K^2 - K_0 = k_0^2 - K_0^2 + 2k_0(k - k_0)$. Substituting these expansions in eq.11 result in,

[12]

$$\psi_t(x,y,t) = \frac{n_t}{2\pi} e^{i(q_{0x} x + q_{0y} y) - \frac{\hbar}{2m}(k_{0x}^2 - K_{0x}^2 + k_{0y}^2 - K_{0y}^2)t} *$$

$$\int A(k_x) A(k_y) e^{i((k_x - k_{0x})(\frac{k_{0x}}{q_{0x}} x - V_x t) + (k_y - k_{0y})(\frac{k_{0y}}{q_{0y}} y - V_y t))} dk_x dk_y$$

where

$$q_x = \sqrt{(k_{0x}^2 - K_{0x}^2)} \qquad q_y = \sqrt{(k_{0y}^2 - K_{0y}^2)}$$

$$V_x = \frac{\hbar k_{0x}}{m} \qquad V_y = \frac{\hbar k_{0y}}{m}$$

Comparing eq.12 to eq.10 ,

$$[13]$$

$$\psi_t(x,y,t) = n' e^{-i\frac{\hbar}{2m}(k_{0x}^2 - K_{0x}^2 + k_{0y}^2 - K_{0y}^2)t} \psi_i(\frac{k_{0x}}{q_{0x}}x - V_x t, \frac{k_{0y}}{q_{0y}}y - V_y t, 0)$$

where n`= $n_t/n_i$. Since Gottfried`s wave packet [8] is not dispersive the time evolution is considered as a phase shift and a shift in the coordinates as seen in eq.13, without changing the wave packet size.

The transmitted wave packet widths in x and y dimensions respectively are,

$$\Delta x_t = \frac{q_{0x}}{k_{0x}}\Delta x_i \quad ; \quad \Delta y_t = \frac{q_{0y}}{k_{0y}}\Delta y_i$$

The variables x and y are independent and from eq.13, we get the relation between the incident wave packet area $\Delta s_i = \Delta x_i \Delta y_i$ and the transmitted wave function area $\Delta s_t = \Delta x_t \Delta y_t$

$$\Delta s_t = \frac{q_{0x}}{k_{0x}}\frac{q_{0y}}{k_{0y}}\Delta x_i \Delta y_i$$

$\Delta s_t$ is the transmitted wave function area, reduced due to the electron loss of kinetic energy.

We take $k_{0x} > k_{0y}$ to have a dominant propagation direction of the free electron along the x direction. Transmitting some of it`s energy to the detector we obtain that electron wave packet size $s_t$ is reduced. This size reduction causes the electron wave function to become too narrow to propagate through two slits and the electron is propagating through only one slit as a result there is no interference pattern.

For explanation of the two slits experiment with a detector type that emits energy signal and by the change in the emitting signal detect the electron location. We look a two paths interferometer experiment [3]. In this experiment an electron wave function propagate through two separate paths,the paths have joint beginning and ending. The interference is determined by the phase difference between in the wave function at the ends of the two paths.When a detector is operating on one of the two paths, the quantum point contact detector`s [10] electron current interacts with the electron wave function in this path. The detector transmits kinetic energy to the measured electron through the coulomb interaction between the electron in the path and the electron current in the detector.This can be obtained from the experimental results that indicate reduction in the detector voltage $V_d$ when the measured electron interact with the detector. Looking at eq .14 we see that decreasing the detector`s electrons effective voltage V decrease the electrons velocity and its kinetic energy.

[14]

$$\beta = \sqrt{1 - (\frac{mc^2}{mc^2 + eV})^2}$$

where m is the electron mass e is the electron charge, V is the effective voltage through the detector, c is the speed of light,β =v/c ,v is the electron velocity. Since the propagation direction of the detector`s electron is the same as the propagation direction of the measured single electron in the path. A momentum transfer from the detector's electrons to the single electron increase it's kinetic energy. Equation.14 is an approximated equation for electron current behavior in solid state. When increasing externally the detector voltage, it is obtains from eq.14 the detector's electrons increase their velocity and kinetic energy and the detector`s electrons transfer to the single electron more kinetic energy.

We write the single electron wave function as a product of the wave function confined at the boundaries potential in the lateral y direction, and a free wave packet moving along the propagation x direction.

$$\psi(x,y,t) = \phi(y,t)\frac{1}{2\pi}\int A(k_x)e^{ik_x x - \frac{\hbar}{2m}k_x^2 t} dk_x$$

The wave function is split into the two interferometer paths:
[15]

$$\psi_1(x,y,t) = A_1\phi(y,t)\frac{1}{2\pi}\int A(k_x)e^{i(k_x x - \frac{\hbar}{2m}k_x^2 t)} dk_x$$

$A_1$, $A_2$ are the probability amplitudes of $\psi_1$ and $\psi_2$.
We looked at the interference along the propagation in x direction.
The increase in the electron kinetic energy is considered as the solution of Schroedinger equation with approximated potential $U = -U_x$ for $x > a$

$$\psi_2(x,y,t) = A_2 \phi(y,t) \frac{1}{2\pi} \int A(k_x) e^{i(k_x x - \frac{\hbar}{2m} k_x^2 t)} dk_x$$

where a is the coordinate at the end of the interaction region between the electron in the path and the detector`s electrons current. Since the electron`s interaction with the detector is only in the small region of the point contact, after the single electron pass through this region we define that the potential $U_x$ equals to the net amount of the observed electron kinetic gain along the propagation direction x.
We consider the electron wave vector main region is around $k_0$
Let $K_0 = (2mv_x/\hbar^2)^{0.5}$, where $k_0 > K_0$. By Taylor series we have,

$$\sqrt{k_x^2 + K_{0x}^2} \approx \sqrt{k_{0x}^2 + K_{0x}^2} + (k_x - k_{0x}) \left[ \frac{d\sqrt{k_x^2 + K_{0x}^2}}{dk_x} \right]_{k=k_0}$$

$$\approx p_0 + (k_x - K_{0x}) \frac{k_{0x}}{p_0}$$

Where $p_0 = (k_0^2 + K_0^2)^{0.5}$. Expanding the time evolution part we get,

$$\psi_t(x,y,t) = \frac{1}{2\pi} A_1 \phi(y) e^{i(p0x - \frac{\hbar}{2m}(k_{0x}^2 + K_{0x}^2)t)} \int A(k_x) e^{i(kx - k0x)(\frac{k0x}{p0x} x - Vt))} dk_x$$

Following the steps we did in the two dimensions free particle we find that the relation between the initial wave function width $\Delta x_i$ and the wave function width after gaining kinetic energy $\Delta x_t$ is,

$$\Delta x_t = \frac{p_0}{k_{0x}} \Delta x_i$$

The increase of the electron`s kinetic energy reduced the main wavevectors region and increase the main electron wave packet width. After the electron wave packet increase due to receiving kinetic energy from the detector`s electrons current there is still interference between the two paths where the wave function in the path with no detector is unchanged. The interference pattern obtain in this case is between two

wave packets with different longitude widths which consisted of two different main regions of wavevectors

The interference equation when considering only the main wavevector value is,

[16]
$$P = A_1^2 + A_2^2 + 2A_1A_2 \cos((k_x - \sqrt{k_x^2 + K_{0x}^2})x - \frac{\hbar K_{0x}^2}{2m}t)$$

Where P is the probability density .The third term in eq.16 is a cosine of the difference between the main wavevectors of the two wave functions, cause reduction of the inference pattern visibility. Summing over all the wavevectors range and not only on the main value and adding wavevector changes in the y axis,will reduce the interference visibility even more. Small increasing in the kinetic energy of the single electron would increase less the wave packet size and the interference loss would be smaller.This is obtained in the experiment [3] results. In this experiment the interference visibility reduced gradually, correlated to increase of the voltage through the detector.This is as we discussed result an increasing of the kinetic energy of the electrons current in the detector,which increase the kinetic energy gained by the electron in the path. Changes in kinetic energy of the particle wave packet doesn`t necessarily cause loss of coherence . In some cases there is a continuation of phase during the energy change process,this appears for example in D.E Pritchard et al, experience [11]. In this experiment a coherent beam of atomic sodium`s scattered by photons. The photons changed the x component of the atoms momentum resulting loss of the interference pattern created by propagating the atomic beam through diffraction grating. When detecting only a narrow part of the scattered beam ,which relate to a certain momentum change most of the interference pattern regain.This indicate that wavevectors change in the particle wavepakect due to kinetic energy gaining, did not cause loss of the wave packet coherence. Then two particles that were coherent before the kinetic energy change are coherent after this energy change. In a strongly disordered environment and noise it is not expected that the particle remain coherent, in these cases an additional effect of dephasing occur [6].

The results of our research can be extended to Schroedinger equation with variable potential that represents a successive discrete or continuos transmission of energy between the single particle and the detector.

In conclusion the detection process causes changes in the wave function size due to energy exchanges between the particle wave function and the detector .In experiment where the particle transmit energy to the detector, the particle wave function, described as a wave packet, reduced it`s distribution uncertainty size. We described these changes in two and three dimensions for area and volume reduction respectively.This size reduction cause loss of inference pattern.The magnitude of the size change is correlated to the amount of kinetic energy lost. We obtain that e particle's kinetic energy gain from a detector oncreasr the particle wave function distribution size. The interference loss in the two paths interferometer experiment explained in addition to the dephasing effect, as an interference between one path in which the wave function is unchanged and a second path which a wave function inside increased it`s wave packet length due to kinetic energy gains from the detector.

The paper was written in plural only as a literary style.     I like to thank G. Kventsel for valuable discussions.